\documentstyle[aps,prl]{revtex}

\begin{document}

\twocolumn

\title{Phase diffusion of Bose-Einstein Condensation close to zero temperature}

\author{Hongwei Xiong,$^{1}$ Shujuan Liu,$^{1}$ Guoxiang Huang$^{2,3}$}
\address{$^{1}$Department of Applied Physics, Zhejiang
University of Technology, Hangzhou, 310032, China}
\address{$^{2}$Department of Physics, East China Normal University, Shanghai, 200062, 
China}
\address{$^{3}$Laboratorie de Physique Th\`{e}orique
de la Mati\`{e}re Condens\`{e}e, case
7020, 2 Place Jussieu, 75251 Paris Cedex 05, France}

\date{\today}
\maketitle

\begin{abstract}
{\it The correlation function  of the quantum fluctuations due to collective excitations
is calculated and used to investigate the phase diffusion of a Bose-Einstein condensate close
to zero temperature.  It is shown that
the phase diffusion time of the condensate is much longer than the result obtained by assuming that
the correlation time of the quantum fluctuations is infinity.
hongweixiong@hotmail.com}
\end{abstract}

{PACS number(s): 03.75.Fi, 05.30.Jp}



\section{Introduction}


The development of the technologies of laser trapping and evaporative
cooling has yielded intriguing Bose-Einstein condensates (BECs) \cite
{ALK,ALK1,ALK2}, a state of matter in which many atoms are in the same
quantum mechanical state. The remarkable observations of gaseous BECs have
opened up new avenues \cite{RMP,RMP1,RMP2} of research into the physical
properties and nature of Bose-condensed systems. The phase properties of a
BEC are of particular interest because the phase of an order parameter, {\it %
i.e.}, the macroscopic wave function of the condensate, reflects directly
the coherent nature of the condensate.

For BEC created in experiment, one of the most important characters is that
all the atoms in the condensate can be described by the wave function ({\it %
i.e.}, the order parameter) with a single phase. Due to thermal and quantum
fluctuations, however, the single phase of the condensate will become
unpredictable beyond the phase diffusion time. After the realization of
BECs, the phase diffusion of the condensate has been discussed intensively%
\cite
{WRIGHT,WRIGHT1,LEW,LEW1,NAR,JAV,MOLMER,GRAHAM,GRAHAM1,LEGGETT,JAK,KUANG,DALVIT,KUKLOV}%
. In particular, the role of quantum fluctuations on the phase diffusion
process was investigated in the pioneering work by Lewenstein and You \cite
{LEW,LEW1}, and a far off-resonant light scattering experiment was proposed
to detect the quantum diffusion. Recently, a Langevin equation was given by
Graham \cite{GRAHAM,GRAHAM1} to discuss the phase diffusion due to quantum
fluctuations and thermal fluctuations. The calculation of the time scale of
the phase diffusion is a very important problem because the phase diffusion
time determines when the phase of a BEC would be unpredictable. Recently,
the phase correlation has been investigated experimentally by the JILA group 
\cite{JILA}. It was found that there is no detectable diffusion of the phase
on time scale $100$ {\rm ms}. The stable interference patterns shown in the
experiment put forward a question \cite{JILA} why the phase correlation is
so robust despite the phase diffusion and complicated rearrangement dynamics
of the two condensates.

In the present work, we address the question of the phase diffusion process
of a condensate close to zero temperature. In general, the phase diffusion
of the condensate can have either a thermal or a quantum origin. At
extremely low temperature (in the experiment by the JILA group \cite{JILA},
the temperature is only $0.1T_{c}$, where $T_{c}$ denotes the critical
temperature of the Bose gas.), the thermal fluctuations can be omitted and
hence the quantum fluctuations become dominant. We give therefore emphasis
on the role of collective excitations due to quantum fluctuations in a phase
diffusion process. Although the phase diffusion process due to quantum
fluctuations has been investigated by several authors such as the recent
researches in \cite{GRAHAM,GRAHAM1,KUKLOV}, the analysis of the time
correlation of the quantum fluctuations is not given when the phase
fluctuations are calculated. Obviously, the correct consideration of the
time correlation of the quantum fluctuations would make more reliable
prediction on the phase diffusion process. In particular, researches show
that the phase diffusion time is much longer than the correlation time of
the quantum fluctuations. In this case, our results show that the phase
diffusion time calculated from the correlation function of the quantum
fluctuations is much longer than that obtained in the previous theoretical
researches \cite{GRAHAM,GRAHAM1,KUKLOV}.

The paper is organized as follows. In Section II, we investigate the phase
fluctuations of the condensate\vspace{1pt} due to quantum fluctuations. In
Sec. III, the phase diffusion time is calculated for the condensate close to
zero temperature, where the effect of thermal fluctuations can be omitted.
Sec. IV contains a discussion and summary of our results.


\section{Phase fluctuations of the condensate due to quantum fluctuations}


For temperature below the critical temperature $T_{c}$, the condensate can
be described very well by the following order parameter with a phase factor $%
\phi \left( t\right) $

\begin{equation}
\Phi \left( {\bf r},t\right) =\Phi _{0}\left( {\bf r}\right) e^{-i\phi
\left( t\right) },  \label{phase1}
\end{equation}
where the phase of the condensate has the form

\begin{equation}
\phi \left( t\right) =\mu \left( N_{{\bf 0}},T\right) t/\hbar,
\label{phase2}
\end{equation}
and the time-independent real component $\Phi _{0}\left( {\bf r}\right) $ is
determined by the stationary Gross-Pitaevskii (GP) equation \cite{RMP}:

\begin{equation}
\mu \left( N_{{\bf 0}},T\right) \Phi _{0}\left( {\bf r}\right) =\left( -%
\frac{\hbar ^{2}}{2m}\nabla ^{2}+V_{{\rm ext}}\left( {\bf r}\right) +g\Phi
_{0}^{2}\left( {\bf r}\right) \right) \Phi _{0}\left( {\bf r}\right) ,
\label{GPequation}
\end{equation}
where $V_{{\rm ext}}\left( {\bf r}\right) $ is an external harmonic
potential, and $g=4\pi \hbar ^{2}a_{s}/m$ is the coupling constant fixed by
the $s-$wave scattering length $a_{s}$. The chemical potential $\mu \left(
N_{{\bf 0}},T\right) $ in the above equation is determined by the
normalization condition for the density distribution $n_{0}\left( {\bf r}%
\right) $ of the condensate. With a Thomas-Fermi approximation \cite{RMP},
one gets easily the following expression for the chemical potential:

\begin{equation}
\mu \left( N_{{\bf 0}},T\right) =\frac{\hbar \omega _{{\rm ho}}}{2}\left( 
\frac{15N_{{\bf 0}}a_{s}}{a_{{\rm ho}}}\right) ^{2/5},  \label{chemical}
\end{equation}
where $\omega _{{\rm ho}}=\left( \omega _{x}\omega _{y}\omega _{z}\right)
^{1/3}$ is the geometric average of oscillator frequencies, and $a_{{\rm ho}%
}=\sqrt{\hbar /m\omega _{{\rm ho}}}$ is the harmonic oscillator length of
the system. From Eqs. (\ref{phase2}) and (\ref{chemical}), we see that the
particle number fluctuations of the condensate yield fluctuations in the
chemical potential, and hence lead to the phase diffusion of the condensate.

Assuming the mean ground state occupation number is $\left\langle N_{{\bf 0}
}\right\rangle $, the average phase of the condensate is then given by

\begin{equation}
\phi \left( \left\langle N_{{\bf 0}}\right\rangle ,t\right) =\mu \left(
\left\langle N_{{\bf 0}}\right\rangle ,T\right) t/\hbar .
\label{averagephase}
\end{equation}
The phase diffusion of the condensate can be described by considering the
phase difference $\Delta \phi \left( t\right) =\phi \left( t\right) -\phi
\left( \left\langle N_{{\bf 0}}\right\rangle ,t\right) $. From Eqs. (\ref
{phase2}) and (\ref{averagephase}), it is straightforward to obtain a
differential equation on \ $\Delta \phi \left( t\right) $:

\begin{equation}
\frac{d\Delta \phi \left( t\right) }{dt}=X_{{\rm qua}}\left( t\right) =\frac{%
\partial \mu \left( \left\langle N_{{\bf 0}}\right\rangle ,T\right) }{%
\partial \left\langle N_{{\bf 0}}\right\rangle }\Delta N_{{\bf 0}}\left(
t\right) /\hbar ,  \label{diff}
\end{equation}
where $X_{{\rm qua}}\left( t\right) $ is determined by collective
excitations due to quantum fluctuations. In the above expression, $\Delta N_{%
{\bf 0}}\left( t\right) $ represents the fluctuations of the ground state
occupation number around $\left\langle N_{{\bf 0}}\right\rangle $. A similar
equation was derived and used by Graham \cite{GRAHAM,GRAHAM1}\ to discuss
the phase diffusion of the condensate. Note that $\Delta N_{{\bf 0}}\left(
t\right) $ can be either negative or positive numbers. For $\Delta N_{{\bf 0}%
}\left( t\right) <0$, there are collective excitations created so that the
atoms would loss in the condensate. Similarly, $\Delta N_{{\bf 0}}\left(
t\right) >0$ means the annihilation of collective excitations, and the
ground state occupation number would increase in this case. In addition, $%
\Delta N_{{\bf 0}}\left( t\right) $ should be time-dependent because it
originates from quantum fluctuations. Eq. (\ref{diff}) is our starting point
to discuss the phase diffusion of the condensate. Because it is obtained
from Eqs. (\ref{phase2}) and (\ref{averagephase}), rather than directly from
a time-dependent GP equation, we anticipate that Eq. (\ref{diff}) is still
correct for longer time where the time-dependent GP equation may be no
longer valid \cite{CASTIN}.

From Eq. (\ref{diff}), the phase fluctuations of the condensate are given by

\begin{equation}
\langle \left( \Delta \phi \left( \tau \right) \right) ^{2}\rangle
=\int_{0}^{\tau }\int_{0}^{\tau }\left\langle X_{{\rm qua}}\left( \xi
\right) X_{{\rm qua}}\left( \xi ^{\prime }\right) \right\rangle d\xi d\xi
^{\prime },  \label{phasewid}
\end{equation}
where $\left\langle X_{{\rm qua}}\left( \xi \right) X_{{\rm qua}}\left( \xi
^{\prime }\right) \right\rangle $ is the correlation function of the quantum
fluctuations. When obtaining Eq. (\ref{phasewid}), we have assumed that $%
\langle \left( \Delta \phi \right) ^{2}\rangle =0$ at time $t=0$. The
calculation of $\left\langle X_{{\rm qua}}\left( \xi \right) X_{{\rm qua}%
}\left( \xi ^{\prime }\right) \right\rangle $ plays a crucial role in
investigating the phase diffusion process close to zero temperature.

For the Bose gas trapped in a harmonic potential, it is convenient to use
the following decomposition of the particle field operator

\begin{equation}
\hat \psi \left( \vec{r}\right) =\Phi \left( \vec{r}\right) +\sum_{i}\left(
u_{i}\left( \vec{r}\right) \alpha _{i}+v_{i}^{\ast }\left( \vec{r}\right)
\alpha _{i}^{\dagger}\right) ,  \label{decom}
\end{equation}
where $\Phi \left( \vec{r}\right) =\left\langle \hat \psi \left( \vec{r}%
\right) \right\rangle $ is the well known order parameter, and the index $i$
labels the elementary excitations of the system. For the collective
excitations discussed here, the energy of the collective mode indexed by $nl$
is given by the dispersion law \cite{STRINGARI}

\begin{equation}
\varepsilon _{{\rm nl}}=\hbar \omega _{{\rm ho}}\left(
2n^{2}+2nl+3n+l\right) ^{1/2}.  \label{energy}
\end{equation}
As shown in \cite{GIO,XIONG1}, the contributions to condensate fluctuations
due to quantum fluctuations are dominated by these phonon-type collective
excitations. For the collective mode $nl$, one obtains the following leading
behaviour for $u_{{\rm nl}}\left( \vec{r}\right) $ and $v_{{\rm nl}}\left( 
\vec{r}\right) $ \cite{GRI}:

\begin{equation}
u_{{\rm nl}}\left( \vec{r}\right) \simeq -v_{{\rm nl}}\left( \vec{r}\right)
\simeq \sqrt{\frac{gn_{0}\left( \vec{r}\right) }{2\varepsilon _{{\rm nl}}}}%
\chi _{{\rm nl}}\left( \vec{r}\right) ,  \label{uvfunction}
\end{equation}
where $\chi _{{\rm nl}}\left( \vec{r}\right) $ is the velocity potential
associated with the collective mode, and satisfies the condition $\int d\vec{%
r}\chi _{{\rm nl}}^{\ast }\left( \vec{r}\right) \chi _{{\rm ij}}\left( \vec{r%
}\right) d^{3}\vec{r}=\delta _{{\rm nl,ij}}$. In addition, the average
occupation number of the atoms corresponding to the collective mode indexed
by $nl$ is given by \cite{BOG,GIO2}

\begin{equation}
\left\langle N_{{\rm nl}}\right\rangle =\left( u_{{\rm nl}}^{2}+v_{{\rm nl}%
}^{2}\right) f_{{\rm nl}},  \label{anumber}
\end{equation}
where $f_{{\rm nl}}=\left[ \exp \left( \varepsilon _{{\rm nl}}/k_{B}T\right)
-1\right] ^{-1}$. When a collective excitation with index $nl$ is created
from the condensate due to quantum fluctuations, its energy $\varepsilon _{%
{\rm nl}}$ originates from the energy fluctuations $\Delta E$ of the
condensate. Under this consideration, a time-energy uncertainty relation can
be used to calculate the longevity $\tau _{{\rm nl}}$ of the collective mode 
$nl$. The longevity $\tau _{{\rm nl}}$ of the collective mode $nl$ is
therefore approximated as $1/\omega _{{\rm ho}}\left( 2n^{2}+2nl+3n+l\right)
^{1/2}$. For JILA experiment \cite{JILA}, this means that the longevity of
the collective mode is smaller than $10$ {\rm ms}, which is obviously much
smaller than the phase diffusion time.

When all collective modes are considered, $X_{{\rm qua}}\left( t\right) $
can be written as:

\begin{equation}
X_{{\rm qua}}\left( t\right) =\frac{\partial \mu \left( \left\langle N_{{\bf %
0}}\right\rangle ,T\right) }{\partial \left\langle N_{{\bf 0}}\right\rangle }%
\sum_{{\rm nl\neq 0}}\Delta N_{{\rm nl}}\left( t\right) /\hbar ,
\label{force}
\end{equation}
where $\Delta N_{{\rm nl}}\left( t\right) $ reflects the changes of the
ground state occupation number due to the creation and annihilation of the
collective mode $nl$. Therefore, the magnitude of $\Delta N_{{\rm nl}}\left(
t\right) $ can be regarded as $\left\langle N_{{\rm nl}}\right\rangle $.
Note that $\Delta N_{{\rm nl}}\left( t\right) $ itself can be either
positive or negative, and varies with time due to quantum fluctuations. In
the case of $\Delta N_{{\rm nl}}\left( t\right) <0$, there are $\left|
\Delta N_{{\rm nl}}\left( t\right) \right| $ atoms created from the
condensate due to quantum fluctuations, while $\Delta N_{{\rm nl}}\left(
t\right) >0$ represents the annihilation of $\Delta N_{{\rm nl}}\left(
t\right) $ atoms.

From Eq. (\ref{force}), $\left\langle X_{{\rm qua}}\left( \xi \right) X_{%
{\rm qua}}\left( \xi ^{\prime }\right) \right\rangle $ can be written as:

$${
\left\langle X_{{\rm qua}}\left( \xi \right) X_{{\rm qua}}\left( \xi
^{\prime }\right) \right\rangle =\frac{1}{\hbar ^{2}}\left[ \partial \mu
\left( \left\langle N_{{\bf 0}}\right\rangle ,T\right) /\partial
\left\langle N_{{\bf 0}}\right\rangle \right] ^{2}\times
}$$

\begin{equation}
 ~~~~~~~~~~~~~~~~~~~~~~~\sum_{{\rm nl\neq 0}}\sum_{%
{\rm n}^{\prime }{\rm l}^{\prime }{\rm \neq 0}}\left\langle \Delta N_{{\rm nl%
}}\left( \xi \right) \Delta N_{{\rm n}^{\prime }{\rm l}^{\prime }}\left( \xi
^{\prime }\right) \right\rangle .  \label{force1}
\end{equation}
Assuming that there is no correlation between different collective modes, 
{\it i.e.}, $\left\langle \Delta N_{{\rm nl}}\left( \xi \right) \Delta N_{%
{\rm n}^{\prime }{\rm l}^{\prime }}\left( \xi ^{\prime }\right)
\right\rangle =0$ when $nl\neq n^{\prime }l^{\prime }$, one gets the
following expression for the correlation function:

$${
\left\langle X_{{\rm qua}}\left( \xi \right) X_{{\rm qua}}\left( \xi
^{\prime }\right) \right\rangle =\frac{1}{\hbar ^{2}}\left[ \partial \mu
\left( \left\langle N_{{\bf 0}}\right\rangle ,T\right) /\partial
\left\langle N_{{\bf 0}}\right\rangle \right] ^{2}\times
}$$

\begin{equation}
  ~~~~~~~~~~~~~~~~~~~~~~~~\sum_{{\rm nl\neq 0}%
}\langle N_{{\rm nl}}\rangle ^{2}e^{-\left| \xi -\xi ^{\prime }\right| /\tau
_{{\rm nl}}}.  \label{froce2}
\end{equation}
When obtaining the above result, we have used the following relation

\begin{equation}
\left\langle \Delta N_{{\rm nl}}\left( \xi \right) \Delta N_{{\rm nl}}\left(
\xi ^{\prime }\right) \right\rangle =\langle N_{{\rm nl}}\rangle
^{2}e^{-\left| \xi -\xi ^{\prime }\right| /\tau _{{\rm nl}}}.
\label{idential}
\end{equation}
In the above expression, $\tau _{{\rm nl}}$ is the longevity of the
collective excitation $nl$. When $\left| \xi -\xi ^{\prime }\right| $ is
much larger than $\tau _{{\rm nl}}$, the correlation between the collective
excitations at times $\xi $ and $\xi ^{\prime }$ can be omitted. Therefore, $%
\tau _{{\rm nl}}$ can be approximated as the correlation time of the
correlation function $\left\langle \Delta N_{{\rm nl}}\left( \xi \right)
\Delta N_{{\rm nl}}\left( \xi ^{\prime }\right) \right\rangle $.

Because the contributions to the quantum fluctuations come mainly from the
low-lying collective excitations, as a reasonable approximation, the
correlation function $\left\langle X_{{\rm qua}}\left( \xi \right) X_{{\rm %
qua}}\left( \xi ^{\prime }\right) \right\rangle $ can be approximated as an
exponential form:

\begin{equation}
\left\langle X_{{\rm qua}}\left( \xi \right) X_{{\rm qua}}\left( \xi
^{\prime }\right) \right\rangle =Q_{{\rm qua}}e^{-\left| \xi -\xi ^{\prime
}\right| /\tau _{{\rm qua}}}.  \label{correlation}
\end{equation}
When the above exponential form is used, $\tau _{{\rm qua}}$ should be
regarded as the average correlation time of the collective excitations, and
is determined by the following expression:

\begin{equation}
\tau _{{\rm qua}}^{2}=\frac{\int_{-\infty }^{\infty }d\tau \tau
^{2}\left\langle X_{{\rm qua}}\left( t\right) X_{{\rm qua}}\left( t+\tau
\right) \right\rangle }{\int_{-\infty }^{\infty }d\tau \left\langle X_{{\rm %
qua}}\left( t\right) X_{{\rm qua}}\left( t+\tau \right) \right\rangle }.
\label{ctime}
\end{equation}
In terms of Eqs. (\ref{froce2}) and (\ref{ctime}), the correlation time of
the quantum fluctuations is given by $\tau _{{\rm qua}}=\sqrt{2}/\omega _{%
{\rm ho}}$. In addition, in Eq. (\ref{correlation}), the magnitude $Q_{{\rm %
qua}}$ of the correlation function is given by

\begin{equation}
Q_{{\rm qua}}=\frac{1}{\hbar ^{2}}\left[ \partial \mu \left( \left\langle N_{%
{\bf 0}}\right\rangle ,T\right) /\partial \left\langle N_{{\bf 0}%
}\right\rangle \right] ^{2}\left\langle \delta ^{2}N_{{\rm qua}%
}\right\rangle .  \label{fmagnitude}
\end{equation}
In the above expression, $\left\langle \delta ^{2}N_{{\rm qua}}\right\rangle
=\sum_{{\rm nl\neq 0}}\langle N_{{\rm nl}}\rangle ^{2}$ can be taken as the
particle number fluctuations \cite{XIONG1} of the condensate due to the
collective excitations. Using the formulas (\ref{energy})-(\ref{anumber}),
after a straightforward (although rather complex) calculation, we obtain the
result of $\left\langle \delta ^{2}N_{{\rm qua}}\right\rangle $:

$${
\left\langle \delta ^{2}N_{{\rm qua}}\right\rangle =0.958\left( \frac{a_{s}}{%
a_{{\rm ho}}}\right) ^{4/5}\left( \frac{T}{T_{c}}\right) ^{2}N^{22/15}
+
}$$

\begin{equation}
 ~~~~~~~~~~~~14.174\left( \frac{a_{s}}{a_{{\rm ho}}}\right) ^{4/5}N^{12/15},
\label{fluc}
\end{equation}
where $N$ is the total number of atoms in the trap. The second term on the
right hand side of the above equation represents the fluctuations due to the
effect of the quantum depletion which is given in \cite{GIO}. This term has
a finite contribution to the condensate fluctuations when the temperature
approaches zero. Thus, we anticipate that there is still phase diffusion in
the case of zero temperature.

We now turn to discussing the phase diffusion of the condensate due to
quantum fluctuations. From Eqs. (\ref{phasewid}) and (\ref{correlation}),
the phase fluctuations of the condensate, which play a crucial role in
discussing the phase diffusion, read

\begin{equation}
\langle \left( \Delta \phi \left( \tau \right) \right) ^{2}\rangle =2Q_{{\rm %
qua}}\tau _{{\rm qua}}\left( \tau -\tau _{{\rm qua}}+\tau _{{\rm qua}%
}e^{-\tau /\tau _{{\rm qua}}}\right) .  \label{fphasewid}
\end{equation}
The phase diffusion time $\tau _{{\rm phase}}$ can be obtained by setting $%
\langle \left( \Delta \phi \left( \tau \right) \right) ^{2}\rangle =\pi ^{2}$
in the above expression.


\section{Phase diffusion time of the condensate close to zero temperature}


We now turn to discussing the phase diffusion time using the phase
fluctuations given by Eq. (\ref{fphasewid}). It is useful to discuss the
phase fluctuations given by Eq. (\ref{fphasewid}) for two special cases.
When the time $\tau $ is much larger than the time scale of the correlation
time $\tau _{{\rm qua}}$, the phase fluctuations of the condensate can be
approximated as:

\begin{equation}
\langle \left( \Delta \phi \left( \tau \right) \right) ^{2}\rangle \approx
2Q_{{\rm qua}}\tau _{{\rm qua}}\tau .  \label{afphasewid}
\end{equation}
Therefore, if the phase diffusion time $\tau _{{\rm phase}}$ calculated from
Eq. (\ref{fphasewid}) is much larger than $\tau _{{\rm qua}}$, the phase
diffusion time in this situation takes the following analytical form: 
\begin{equation}
\tau _{{\rm phase}}=\frac{\pi ^{2}}{2Q_{{\rm qua}}\tau _{{\rm qua}}}.
\label{phasediff}
\end{equation}

In the case of $\tau <<\tau _{{\rm qua}}$, however, the phase fluctuations
of the condensate can be approximated as:

\begin{equation}
\langle \left( \Delta \phi \left( \tau \right) \right) ^{2}\rangle \approx
Q_{{\rm qua}}\tau ^{2}.  \label{usualphase}
\end{equation}
Different from the result given by Eq. (\ref{afphasewid}), the phase
fluctuations are proportional to $\tau ^{2}$ when $\tau $ is much smaller
than the correlation time $\tau _{{\rm qua}}$. Therefore, if $\tau _{{\rm %
phase}}$ calculated from Eq. (\ref{fphasewid}) is much smaller than $\tau _{%
{\rm qua}}$, the analytical result of the phase diffusion time $\tau _{{\rm %
phase}}^{\prime }$ is then

\begin{equation}
\tau _{{\rm phase}}^{\prime }=\frac{\pi \hbar }{\delta N_{{\rm qua}}\partial
\mu \left( \left\langle N_{{\bf 0}}\right\rangle ,T\right) /\partial
\left\langle N_{{\bf 0}}\right\rangle },  \label{uphasediff}
\end{equation}
where $\delta N_{{\rm qua}}=\sqrt{\left\langle \delta ^{2}N_{{\rm qua}%
}\right\rangle }$. 

We now turn to discuss the phase correlation experiment by the JILA group 
\cite{JILA}. The experimental values in the experiment are: $N_{{\bf 0}%
}=5\times 10^{5}$, $T\approx 50$ {\rm nk}, $T_{c}\approx 500$ {\rm nk}, and $%
a_{s}\approx 5\times 10^{-7}$ {\rm cm}. It may be helpful to make a
comparison between the particle number fluctuations due to quantum
fluctuations and thermal fluctuations. For temperature much lower than the
critical temperature, the analytical result $\left\langle \delta ^{2}N_{{\rm %
th}}\right\rangle =\pi ^{2}N\left( T/T_{c}\right) ^{3}/6\zeta \left(
3\right) $ \cite{XIONG1} can give a rather well description for the particle
number fluctuations due to thermal fluctuations. For the values typical for
the experiment by the JILA group \cite{JILA}, a simple calculation shows
that $\left\langle \delta ^{2}N_{{\rm qua}}\right\rangle /\left\langle
\delta ^{2}N_{{\rm th}}\right\rangle =62.6$. Therefore, the thermal
fluctuations can be safely omitted when the phase diffusion process is
investigated for the experiment by JILA group \cite{JILA}. Using the formula
(\ref{fphasewid}) (or Eq. (\ref{phasediff})), the numerical result of the
phase diffusion time $\tau _{{\rm phase}}$ is $119$ {\rm s}, which is much
larger than the correlation time $\tau _{{\rm qua}}$. When obtaining this
result, we have used the exponential form of the correlation function given
by Eq. (\ref{correlation}). In fact, we can obtain $\tau _{{\rm phase}}$
directly from Eqs. (\ref{phasewid}) and (\ref{froce2}), and it is worth
pointing out that there is no important correction to the phase diffusion
time, in comparison with the result obtained by using the exponential form (%
\ref{correlation}). The merit of the exponential form (\ref{correlation}) is
that it clearly shows the role of particle number fluctuations on the phase
diffusion process, and the analytical result of the phase fluctuations is
rather concise using this exponential form.

If the correlation time of the collective excitations is assumed to be
infinity, however, using Eqs. (\ref{fluc}) and (\ref{uphasediff}), the
numerical result of $\tau _{{\rm phase}}^{\prime }$ is $0.62$ {\rm s}, which
is much smaller than the result given by Eq. (\ref{phasediff}). Although the
phase fluctuations due to quantum fluctuations are investigated in deep in
Ref. \cite{KUKLOV}, the finiteness of the correlation time of the quantum
fluctuations was not considered, and the dephasing time was approximated as $%
1$ {\rm s}. In addition, it is worth pointing out that although a Langevin
equation was proposed by Graham \cite{GRAHAM,GRAHAM1} to investigate the
phase fluctuations due to thermal fluctuations and quantum fluctuations, the
phase fluctuations due to collective excitations were proportional to $\tau
^{2}$, because the finiteness of the longevity of the collective excitations
was not considered too.

For temperature close to zero, our result of the phase diffusion time given
by Eq. (\ref{phasediff}) is reasonable because of two reasons: (i) In the
experiment by the JILA group \cite{JILA}, the phase of the condensate was
found to be very robust. In fact, the rigidity of the phase was also shown
in other experiments, such as the observation of the interference between
two BECs \cite{INTER}, and the recent experiments where the optical lattice 
\cite{LATTICE1,LATTICE} is used to investigate the coherent properties of
the BECs. For example, recently a BEC \cite{LATTICE} is created with up to $%
2\times 10^{5}$ atoms and no discernible thermal component. The radial
trapping frequencies are relaxed over a period of $500$ {\rm ms }to $24$ Hz
such that the harmonic potential becomes spherically symmetric. Then three
optical standing waves are aligned orthogonal to each other, in order to
form a three-dimensional lattice potential. In this situation, the
condensate is distributed over more than $150,000$ lattice sites. When the
magnetic trap and lattice potential are both switched off, it is interested
to find that there is a high-contrast interference pattern, which means that
phase is still robust after the BEC has been formed for nearly $1${\rm \ s},
and even after the interference between a large number of BECs. (ii) In the
present work, the correlation time of the quantum fluctuations is calculated
and found to be much smaller than the time scale of the phase diffusion
time. In this situation, we should regard the quantum fluctuations as a
white noise to investigate the phase diffusion process. Recall that the
particular collective excitations are rather stable when it is created in
the experiment \cite{COLLECTIVE,COLLECTIVE1,COLLECTIVE2} by applying a small
time-dependent perturbation, it seems the longevity of the collective
excitations is very long. However, we should note that in the problem
discussed here for the mechanism of the phase diffusion, the collective
excitations are created and annihilated through the quantum fluctuations. As
pointed out in this paper, the longevity of these collective excitations is
found to be much smaller than the phase diffusion time.

We should note that when obtaining the phase diffusion time, the time-energy
uncertainty relation is used to calculate approximately the longevity of the
collective modes created due to quantum fluctuations. A more accurate
average longevity of these collective modes can be obtained when
interparticle interaction effect is included. Nevertheless, a reasonable
order of magnitude on the phase diffusion time can be obtained, using Eq. (%
\ref{fphasewid}) and the time-energy uncertainty relation. Additionally, the
collective time $\tau _{{\rm qua}}$ can be obtained from Eq. (\ref{fphasewid}%
) when $\tau _{{\rm phase}}$ is measured in experiment. This gives us a
chance to check whether standard many-body theory can be used successfully
to investigate the interested quantity $\tau _{{\rm qua}}$.


\section{Discussion and conclusion}


In summary, the phase diffusion time of the condensate due to quantum
fluctuations is discussed at extremely low temperature. Because the
correlation time of the quantum fluctuations is much smaller than the time
scale of the phase diffusion, the quantum fluctuations can be regarded as
white noise when the phase diffusion of the condensate is investigated for
extremely low temperature. In this situation, the phase diffusion time
calculated here is much longer than the result obtained in the previous
theoretical researches \cite{GRAHAM,GRAHAM1,KUKLOV}. It is obvious that the
present work can not be applied directly to the experiment conducted by the
JILA group \cite{JILA}, where the complicated rearrangement of the two
condensates would be very important to the phase diffusion process. We shall
extend our idea in a future work to the case of the specific situation
realized by the JILA group \cite{JILA} and discuss the phase diffusion at
finite temperature. Recently, BECs have been realized in quasi-one and
quasi-two dimensions \cite{EXPLOW}, where new phenomena such as
quasicondensates with a fluctuating phase \cite{PETROV1,PETROV2,KAGAN} may
be observed. A simple method is developed recently \cite
{XIONG1,XIONG2,XIONG21,XIONG22} to discuss the particle number fluctuations
of the low-dimensional condensate. This makes it possible to discuss the
phase diffusion process in low-dimensional condensates.


\section*{Acknowledgments}


This work was supported by Natural Science Foundation of China. One of us
(G. H.) is indebted to National Natural Science Foundation of China, and the
French Ministry of Research for a visiting grant at Universit\`{e} Paris 7.

\end{document}